\documentclass[twocolumn]{article}

\usepackage[scale=0.8]{geometry}
\usepackage[acronym]{glossaries}
\usepackage{subcaption}
\usepackage{tikz}
\usepackage[affil-it]{authblk}
\usepackage{hyperref}
\usepackage{mathpazo}
\usepackage{units}

\setacronymstyle{short-long}

\hypersetup {
    colorlinks=true,
    linkcolor=red,
    filecolor=magenta,      
    urlcolor=cyan,
    pdfpagemode=FullScreen,
}

\title{\textbf{\huge Resolution Studies for Future DIRC Detectors}}

\author{Mustafa Schmidt${}^{1,*}$, Afaf Wasili${}^2$}

\date{\footnotesize ${}^1$University of Wuppertal, Gaussstr. 20, 42119 Wuppertal, Germany \\\footnotesize ${}^2$Department of Physical Sciences, Physics Division, College of Science, Jazan University, P.O. Box. 114, Jazan 45142, Kingdom of Saudi Arabia \\${}^*$Corresponding author: \url{muschmidt@uni-wuppertal.de}}

\newacronym{bnl}{BNL}{Brookhaven National Laboratory}
\newacronym{bsdf}{BSDF}{Bidirectional Scattering Distribution Function}
\newacronym{edd}{EDD}{Endcap Disc DIRC}
\newacronym{dirc}{DIRC}{Detection of Internally Reflected Cherenkov light}
\newacronym{epic}{ePIC}{Electron-Proton/Ion Collider}
\newacronym{fair}{FAIR}{Facility of Antiproton and Ion Research}
\newacronym{mcp}{MCP}{Micro Channel Plate}
\newacronym{panda}{PANDA}{antiProton ANnihilation at DArmstadt}
\newacronym{pmt}{PMT}{Photo Multiplier Tube}
\newacronym{pid}{PID}{Particle IDentification}
\newacronym{rich}{RICH}{Ring-Imaging CHerenkov}
\newacronym{rms}{RMS}{Root Mean Square}
\newacronym{sipm}{SiPM}{Silicon Photomultipliers}
\newacronym{sctf}{SCTF}{Super Charm-Tau Factory}
\newacronym{tof}{TOF}{Time Of Flight}
\newacronym{top}{TOP}{Time Of Propagation}

\begin{document}
\twocolumn[
    \begin{@twocolumnfalse}
        \maketitle
        \begin{abstract}
            \textit{The resolution of Cherenkov detectors using the \gls*{dirc} method is mainly limited by three components: photon loss due to steep angles of incidence, dispersion effects, and angle straggling due to the Coulomb interaction of the charged particle.
            This paper highlights the general limitations of any generic \gls*{dirc} detector based on studies for the proposed experiments \gls*{panda} in Darmstadt and \gls*{sctf} in Russia.
            These studies used theoretical calculations and simplified Monte-Carlo simulations to obtain an upper limit of all individual resolution parameters.
            Therefore, they can be used for feasibility studies in future detector developments.}
        \end{abstract}
    \end{@twocolumnfalse}
    \vspace{2\baselineskip}
]

\tableofcontents

\section{Introduction}

\begin{figure}[!h]
    \centering
    \begin{subfigure}{\linewidth}
        \includegraphics[width=\linewidth]{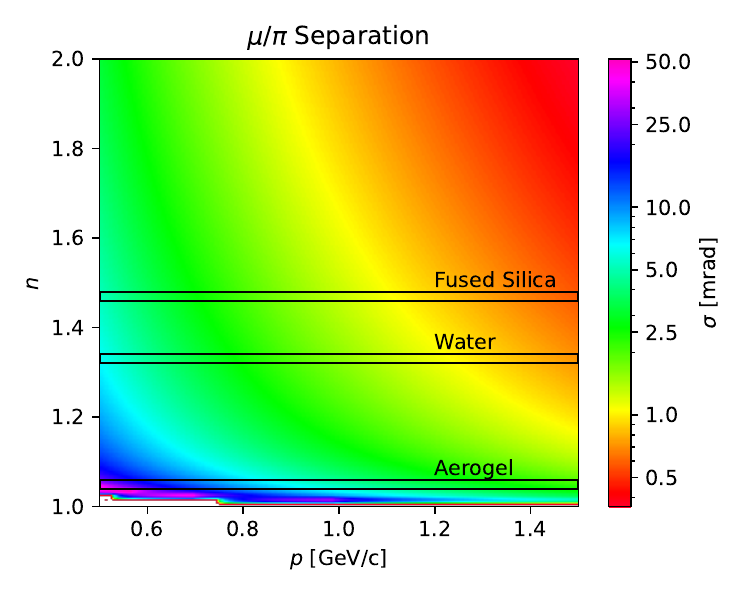}
        \caption{$\mu/\pi$ separation.}
        \label{subfig:required_resolution_mupi}
    \end{subfigure}
    \begin{subfigure}{\linewidth}
        \includegraphics[width=\linewidth]{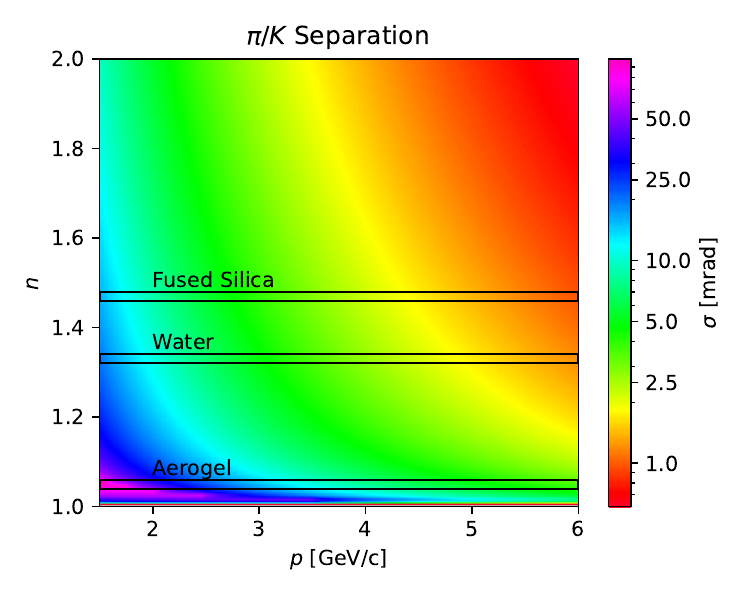}
        \caption{$\pi/K$ separation.}
        \label{subfig:required_resolution_piK}
    \end{subfigure}
    \caption{Required spatial resolution of a DIRC detector for separating two particle species.}
    \label{fig:required_resolution}
\end{figure}

Modern particle detectors usually require excellent \gls*{pid} using \gls*{tof} and Cherenkov counters.
For the future experiments \gls*{panda} \cite{Belias2023OverviewFAIR} at \gls*{fair} in Darmstadt and \gls*{sctf} \cite{Piminov2018ProjectBINP}, which was initially planned to be built in Russia, as well as the planned \gls*{epic} detector \cite{Allaire2024ArtificialAI4EIC} at \gls*{bnl}, several state-of-the-art \gls*{pid} detectors were foreseen.
For \gls*{panda}, the target requirement is the separation of $\pi^\pm$ and $K^\pm$ up to particle momenta of $\unit[4]{GeV/c}$ with a separation power of $n_\sigma = 3\sigma$.
For this purpose, a \gls*{tof} and \gls*{dirc} detector \cite{Singh2019TechnicalDetector}, based on the detector used at BaBar, were planned to be constructed in the barrel region.
For the forward endcap of the target spectrometer, a novel \gls*{dirc} detector called \gls*{edd} \cite{Davi2022TechnicalDIRC} covering smaller polar angles was proposed.
Both \gls*{dirc} counters were designed using fused silica as a radiator and \gls*{mcp}-\glspl*{pmt} for the photon detection due to improved radiation hardness and easier handling without the requirement of any additional cooling for single-photon detection compared to \glspl*{sipm} \cite{Miehling2023LifetimePhotomultipliers}.

The \gls*{sctf} experiment decided on a competition between \gls*{rich} and \gls*{dirc} detectors to find the optimal solution regarding cost-efficiency and feasibility.
The primary focus was separating $\mu^\pm$ and $\pi^\pm$ at low momenta between 0.5 and $\unit[1.5]{GeV}$.
For the \gls*{rich} radiator, aerogel was proposed because of its low refractive index $n$, whereas the design of the proposed \gls*{dirc} detectors was close to the ones for \gls*{panda}.
The significant drawbacks of an aerogel \gls*{rich} detector are the difficulty of producing a large detector covering $4\pi$, the possibility of focusing the Cherenkov light with different aerogel layers, and the massive number of sensors with many readout channels.
Another disadvantage of a \gls*{rich} detector with medium having a low refractive index is the smaller photon yield.
Despite these issues and the increased photon scattering inside aerogel, detailed simulation and test beam studies have shown that the desired separation power for the \gls*{rich} detectors was reachable \cite{Barnyakov2020FARICHResults}.

In the case of \gls*{epic}, a separation of $\pi^\pm$ and $K^\pm$ is desired, which is similar to \gls*{panda} but up to larger momenta around 6\,GeV/c.
This goal is supposed to be achieved by using a combination of \gls*{tof}, \gls*{dirc}, and \gls*{rich} detectors to cover various parts of the phase space.
Both \gls*{sctf} \cite{Barnyakov2023CalibrationFactory, Hayrapetyan2022PlansNovosibirsk} and \gls*{epic} require sophisticated \gls*{dirc} detector designs and precise studies to fulfill the desired specifications \cite{Kalicy2024TheEIC}.
Although \gls*{sctf} will most likely never be built, future experiments might benefit from the resolution studies presented in this paper to determine whether constructing a \gls*{dirc} detector is feasible for the targeted phase space, especially for those where a separation of muons and pions is required.

\section{Required Detector Resolution}

\subsection{Spatial Resolution}

Cherenkov light is emitted if a charged particle passes through a medium with the refractive index $n$ and has a speed greater than the speed of light $c$ in that medium.
The following well-known equation can be used to calculate the opening angle $\theta_c$ of the Cherenkov cone:
\begin{equation}
    \cos\theta_c = \frac{1}{n\beta}
    \label{eq:cherenkov}
\end{equation}
This formula can be rewritten to
\begin{equation}
    \cos\theta_c = \frac{\sqrt{p^2 + m^2}}{np}
    \label{eq:cherenkov_angle}
\end{equation}  
by using $\beta = p/E$ and the energy-momentum relation $E^2 = p^2 + m^2$ of the special theory of relativity.
Now, the Cherenkov angle becomes a function of the particle's mass $m$ and momentum $p$, which underlines the possibility of \gls*{pid} by determining the momentum with external tracking detectors and the Cherenkov angle with a \gls*{dirc} or \gls*{rich} detector.

The required detector resolution of any Cherenkov detector can be directly calculated from equation~(\ref{eq:cherenkov_angle}) by computing the difference of the Cherenkov angles of two particle species:
\begin{eqnarray}
    \sigma = \frac{1}{n_\sigma}\left[\arccos\left(\frac{\sqrt{p^2 + m_2^2}}{np}\right)-\right.\\\left.\arccos\left(\frac{\sqrt{p^2 + m_1^2}}{np}\right)\right]
\end{eqnarray}
In general, a separation power $n_\sigma$ of at least $n_\sigma=3\sigma$ should be targeted as a minimum goal for \gls*{pid} detectors, which is equivalent to a misidentification of 0.13\%.
Hence, this value is also used for the following studies.

Figure~\ref{fig:required_resolution} shows the required detector resolution of a Cherenkov detector as a function of the primary particle's momentum and the nominal refractive index of the material.
These plots reveal that the optimizations for materials with low refractive indices $n$, such as aerogel or water, require less work than those for fused silica or comparable materials.
While the required resolution for $\mu/\pi$ separation at a momentum of 1.5\,GeV/c is located around 2.5\,mrad for aerogel, it must be lower than 0.7\,mrad for fused silica, as visible in Figure~\ref{subfig:required_resolution_mupi}.
For $\pi/K$ separation, shown in Figure~\ref{subfig:required_resolution_piK}, the required resolution is greater because of the larger mass difference between the two particle species.
However, due to its small refractive index, aerogel and materials with similar $n$ are generally used in \gls*{rich} detectors.

\subsection{Time Resolution}

Furthermore, the required time resolution for \gls*{top} counters or additional time-based reconstructions can be calculated.
The propagation time of each photon is given as
\begin{equation}
    t = \frac{s}{v(\lambda)}
    \label{eq:time}
\end{equation}
where $s$ is the travelled distance of the photon and $v$ the group velocity, which is a function of the wavelength:
\begin{equation}
    v(\lambda) = \frac{c}{n-\lambda_0 \frac{\partial n}{\partial \lambda}}
\end{equation}
For these studies, an average wavelength of 500\,nm was assumed.

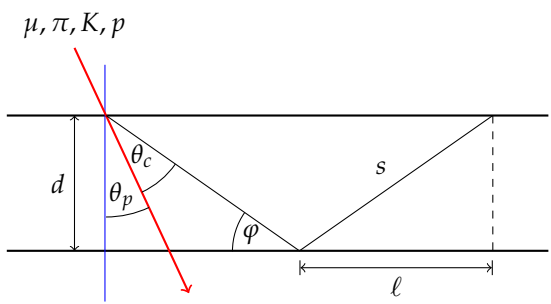
\begin{figure}[!h]
    \centering
    \begin{tikzpicture}[scale=0.9]
        \def\thetap{-65}
        \def\thetac{35}
        \def\xp{1}
        \def\yp{3}
        \def\d{2}
        \def\s{\d/sin(\thetac)}
        \coordinate (intersection) at ({\xp+(2-\yp)/sin(\thetap)*cos(\thetap)},2);
        \draw[thick] (0,0) -- (8,0);
        \draw[thick] (0,2) -- (8,2);
        \draw[blue] (1.45,-0.75) -- (1.45,2.75);
        \draw[thick, red, ->] (1,3) -- ++(\thetap:4);
        \draw (intersection) -- ++(-\thetac:{\s}) node (down) {} -- ++(\thetac:{\s}) node (up) {} node[midway, above left] {$s$};
        \draw (intersection) ++(-65:1.5) arc (-65:-90:1.5);
        \draw (intersection) ++(-\thetac:1.25) arc (-\thetac:-65:1.25);
        \draw (intersection) ++(-\thetac:3.5) ++(145:1) arc (145:180:1);
        \draw (\xp,\yp) node[above] {$\mu,\pi,K,p$};
        \draw (1.7,0.8) node {$\theta_p$};
        \draw (2,1.4) node {$\theta_c$};
        \draw (3.6,0.25) node {$\varphi$};
        \draw[<->] (1,0) -- (1,\d) node[midway, left] {$d$};
        \draw[dashed] (up.center) -- ++(0,-\d) ++(0,-0.25) node (right) {};
        \draw[|<->|] (down.center) ++(0,-0.25) -- (right.center) node[midway, below] {$\ell$};
    \end{tikzpicture}
    \caption{Sketch showing all parameters important to describe the photon path in a \gls*{dirc} detector with a radiator thickness $d$. The primary particle is drawn in red.}
    \label{fig:sketch_dirc}
\end{figure}

\begin{figure}[!h]
    \centering
    \begin{subfigure}{\linewidth}
        \includegraphics[width=\linewidth]{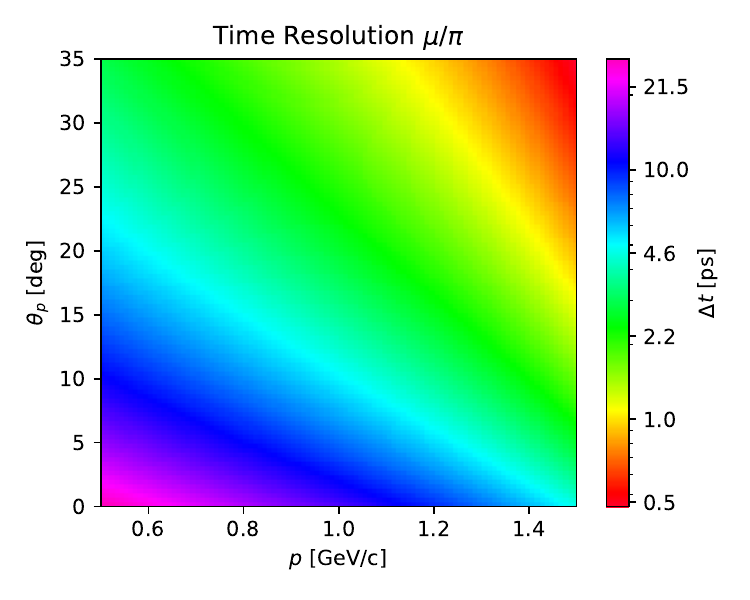}
        \caption{$\mu/\pi$ separation.}
        \label{subfig:time_resolution_mupi}
    \end{subfigure}
    \begin{subfigure}{\linewidth}
        \includegraphics[width=\linewidth]{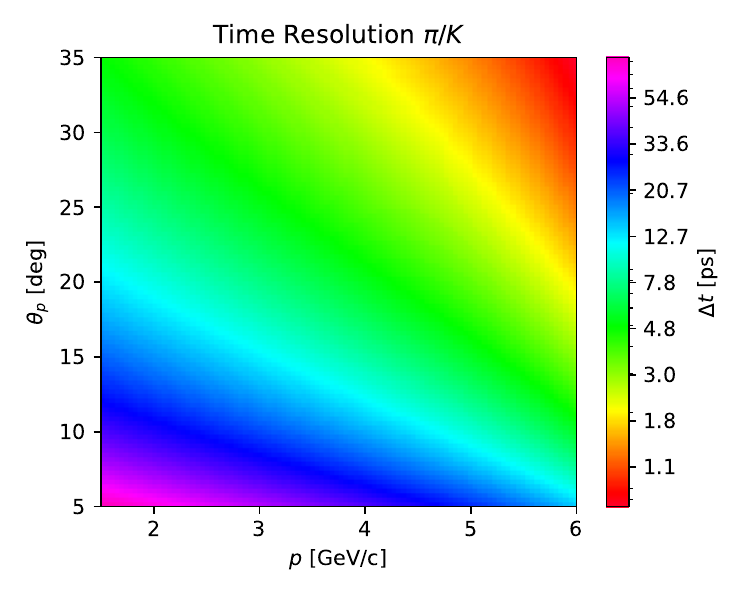}
        \caption{$\pi/K$ separation.}
        \label{subfig:time_resolution_piK}
    \end{subfigure}
    \caption{Required time resolution of a DIRC detector for separating two particle species in a fused silica radiator.}
    \label{fig:time_resolution}
\end{figure}

All essential parameters required to describe the photon path inside a \gls*{dirc} detector are illustrated in Figure~\ref{fig:sketch_dirc}.
The primary particle displayed in red passes through a radiator with the thickness $d$ and emits Cherenkov photons.
The particle's angle is labelled $\theta_p$ and the Cherenkov angle $\theta_c$.
The photon displacement $s$ directly correlates to the travelled distance $\ell$:
\begin{equation}
    s = \frac{\ell}{\cos\varphi}
    \label{eq:distance}
\end{equation}
Here, $\varphi = \pi/2 - \theta$ is the angle between the photon and the radiator surface.
To simplify the following calculations, each Cherenkov photon's total angle of incidence $\theta$ is assumed to be the sum of the particle and photon angles $\theta_p + \theta_c$ and thus identical to photons hitting the sensor plane perpendicular.
The results $\mu/\pi$ separation can be found in Figure~\ref{subfig:time_resolution_mupi} and for $\pi/K$ separation in Figure~\ref{subfig:time_resolution_piK}.
In both cases, the radiator's length was defined as 1\,m.
For $\mu/\pi$ at 1.5\,GeV/c and a particle angle of $10^\circ$, a total time resolution of around 2.2\,ps has to be reached.

\subsection{Resolution Components}

The reconstructed Cherenkov angle would be a $\delta$ distribution at the truth angle for an ideal detector.
However, various smearing effects result in a measured Cherenkov angle distribution ideally centered around the real angle, resulting in a finite detector resolution.
The resolution of a Cherenkov detector can be broken down to the following square-sum of individual resolution components:
\begin{equation}
    \sigma^2 = \frac{\sigma_\mathrm{disp}^2 + \sigma_\mathrm{geo}^2 + \sigma_\mathrm{sens}^2 + \sigma_\mathrm{scat}^2}{N} + \sigma_\mathrm{track}^2 + \sigma_\mathrm{strag}^2 + \sigma_\mathrm{loss}^2
\end{equation}
In this equation, the possible correlation between different terms and higher-order corrections has been ignored.
The parameters stated above represent the following terms:
\begin{itemize}
    \item[$\sigma_\mathrm{disp}$]: Photon dispersion inside the Cherenkov medium resulting in different photon wavelengths
    \item[$\sigma_\mathrm{geo}$]: Effects of the detector geometry, such as focusing optics or expansion volumes
    \item[$\sigma_\mathrm{sens}$]: Granularity of the photon sensors
    \item[$\sigma_\mathrm{scat}$]: Photon scattering inside the material and due to an existing surface roughness
    \item[$\sigma_\mathrm{track}$]: Position and momentum resolution of the tracking detectors
    \item[$\sigma_\mathrm{strag}$]: Angle straggling of the primary particle inside the radiator material
    \item[$\sigma_\mathrm{loss}$]: Energy loss of the primary particle passing the radiator volume
\end{itemize}

In principle, the limitations of the photon sensors' structure and the detector geometry can be optimized without any limitations.
An excellent polishing of the radiator surface removes this smearing part as well.
Hence, the main limitations are given by dispersion effects resulting from photon creation with different wavelengths and the related Cherenkov angle smearing.
All these photon-related errors are purely statistical and therefore scale with the square root $\sqrt{N}$ of the number of detected photons.

The other three components are systematic, as they are related to the primary particle and cannot be mitigated by increasing the number of photons.
Installing extremely high momentum and spatial precision detectors in the final experiment can reduce the tracking error.
As a result, the angle straggling part induces the most significant Cherenkov smearing, especially for low-momentum particles below $p=1.5$\,GeV.

\section{Photon Losses}

\begin{figure}[!h]
    \centering
    \begin{subfigure}{\linewidth}
        \includegraphics[width=0.9\textwidth]{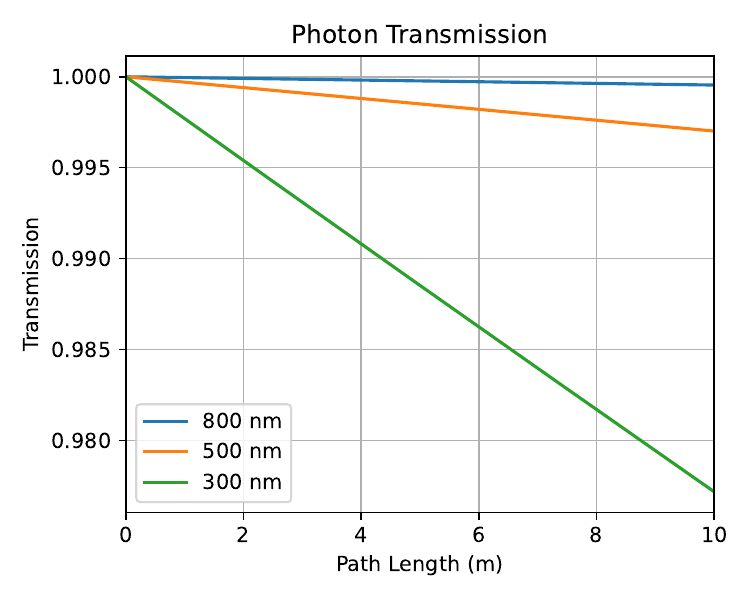}
        \caption{Rayleigh scattering of photons passing a fused silica radiator.}
        \label{subfig:transission}
    \end{subfigure}
    \begin{subfigure}{\linewidth}
        \includegraphics[width=\textwidth]{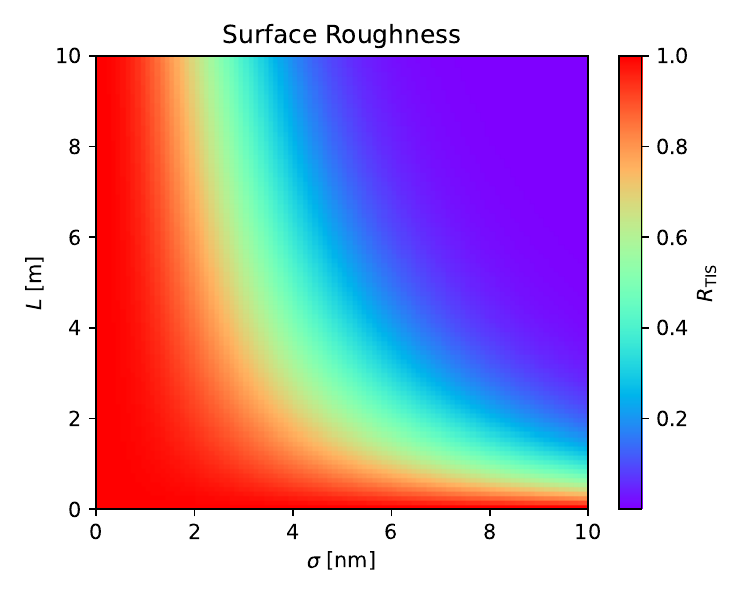}
        \caption{Surface losses of Cherenkov photons of a pion with 1\,GeV/c momentum for a radiator thickness of 2\,cm and $15^\circ$ particle angle.}
        \label{subfig:roughness}
    \end{subfigure}
    \caption{Photon losses due to Rayleigh scattering and surface roughness.}
    \label{fig:photon_transmission}
\end{figure}

Photons in a \gls*{dirc} detector can get lost due to the following reasons:
\begin{enumerate}
    \item Photons leaving the radiator when internal reflection is not fulfilled.
    \item Surface losses due to an existing surface roughness.
    \item Scattering inside the radiator material.
    \item Bulk absorption in the radiator medium due to impurities or missing transmittance for a given wavelength.
\end{enumerate}
\subsection{Bulk Scattering}
Scattering inside the radiator for photons in the visible part of the spectrum can be described with the Rayleigh scattering cross section \cite{Wu2008TextbookI}, which is usually stated as
\begin{equation}
    \sigma_\text{s} (\lambda) = \frac{ 8 \pi}{3}  \left( \frac{2\pi}{\lambda}\right)^4 \left( \frac{ n^2-1}{ n^2+2 } \right)^2 r^6
\end{equation}
with $r$ being the radius of the object scattering the light.
This equation can be transformed into an absorption coefficient  
\begin{equation}
    \alpha_\text{s} (\lambda) = \frac{8 \pi^3}{3 \lambda^4} n^8 p^2 k_B T_\text{f} \beta    
\end{equation}
where $p$ is the photoelastic coefficient of fused silica, $k_B$ the Boltzmann constant, $\beta_T$ the isothermal compressibility, and $T_f$ a fictive temperature.
This coefficient can then be used in combination with the Beer-Lambert law:
\begin{equation}
    I = I_0 e^{-\alpha L}
    \label{eq:photon_losses}
\end{equation}  
to compute the remaining intensity of the light after passing a distance $L$.
Since the absorption coefficient scales with the inverse photon wavelength to the power of four, light with shorter wavelengths is significantly more likely to be scattered.
Realistic results can be obtained by choosing the temperature $T\approx 300$\,K, a photoelastic coefficient $p\approx 0.22$ \cite{Born2000PrinciplesWolf}, and $\beta_T\approx 7\cdot 10^{-11}$\,Pa${}^{-1}$ \cite{Wray1969RefractiveTemperature}.
Assuming a fused silica radiator with a length of 10\,m, the transmission
\begin{equation}
    T = \frac{I}{I_0}
\end{equation}
is plotted in Figure~\ref{subfig:transission}.
Even for blue light around 300\,nm, the losses are around 2\% and can therefore be ignored.

\subsection{Surface Scattering}

The surface losses are directly correlated with the surface roughness.
Integrated the \gls*{bsdf} \cite{Bartell1981titleTheBTDF/title} results in the  following formula
\begin{equation}
    T_\mathrm{TIS} = \left(4\pi \cos\theta R_q \frac{n}{\lambda}\right)^2
\end{equation}
which is equal to the transmission probability of a single photon leaving the radiator via its surface if the surface roughness is small compared to the photon wavelength.
In this equation, $R_q$ is the \gls*{rms} value of the surface roughness.
The probability of remaining inside the radiator after one reflection is then given as
\begin{equation}
    R_\mathrm{TIS} = 1 - T_\mathrm{TIS}
\end{equation}
Hence, the probability for $N$ reflections decreases to
\begin{equation}
    P_N = R_\mathrm{TIS}^N
\end{equation}
The number of reflections is given as
\begin{equation}
    N = \frac{L}{d} \tan\theta
\end{equation}
In Figure~\ref{subfig:roughness}, the reflection probability is plotted as a function of the radiator length $L$ and the \gls*{rms} value of the surface roughness.

The refraction angle $\theta_2$ of a light ray with the angle $\theta_1$ passing from one volume with the refractive index $n_1$ to another one with $n_2$ can be calculated using Snell's law \cite{Kovalenko2001Descartes-SnellAbsorption}:
\begin{equation}
    n_1 \sin(\theta_1) = n_2 \sin(\theta_2)
    \label{eq:snells_law}
\end{equation}
The refractive index of air can be assumed as 1, very close to that of a vacuum.
Internal reflection occurs when the angle $\beta$ is identical to $\pi/2$, i.e., the sine term on the right side vanishes.
This results in
\begin{equation}
    \alpha = \arcsin\left(\frac{1}{n}\right)
    \label{eq:internal_reflection}
\end{equation}
where $n$ is the refractive index of the Cherenkov radiator.

\begin{figure}
    \centering
    \includegraphics[width=\linewidth]{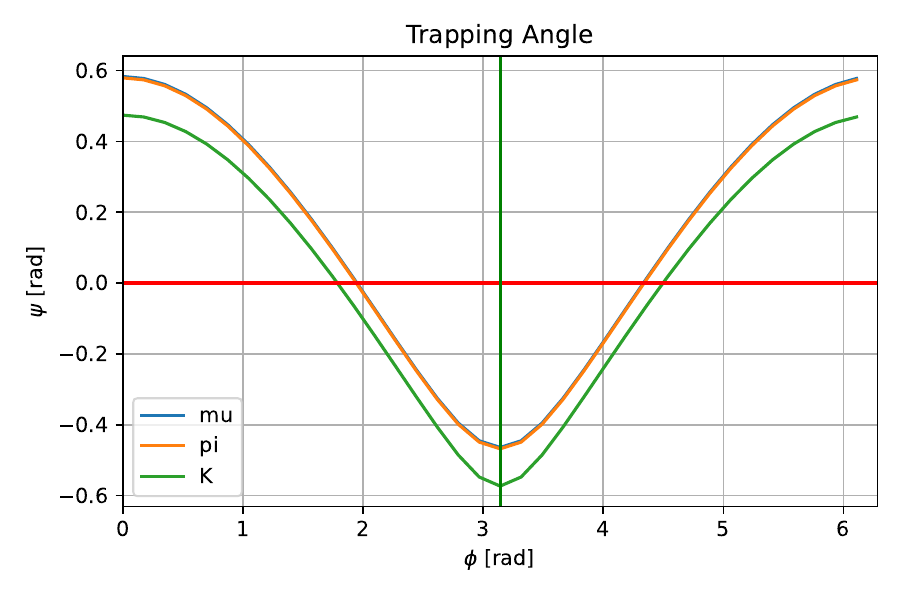}
    \caption{Condition for photon trapping in the radiator.}
    \label{fig:psi}
\end{figure}

\subsection{Trapped Photons}

A \gls*{dirc} detector can only detect photons that fulfill this internal reflection condition, whereas the others are lost.
With equation~(\ref{eq:cherenkov}) it follows:
\begin{equation}
    \arcsin\left(\frac{1}{n}\right) = \arccos\left(\frac{1}{n\beta}\right)
    \label{eq:rindex}
\end{equation}
For $\beta\rightarrow 1$ particles, it immediately follows for the required refractive index
\begin{equation}
    n > \sqrt{2} \approx 1.41
\end{equation}
which is very close to that of fused silica ($n=1.47$), making it an ideal radiator material for \gls*{dirc} detectors.

\begin{figure}[!h]
    \centering
    \begin{subfigure}{\linewidth}
        \includegraphics[width=\textwidth]{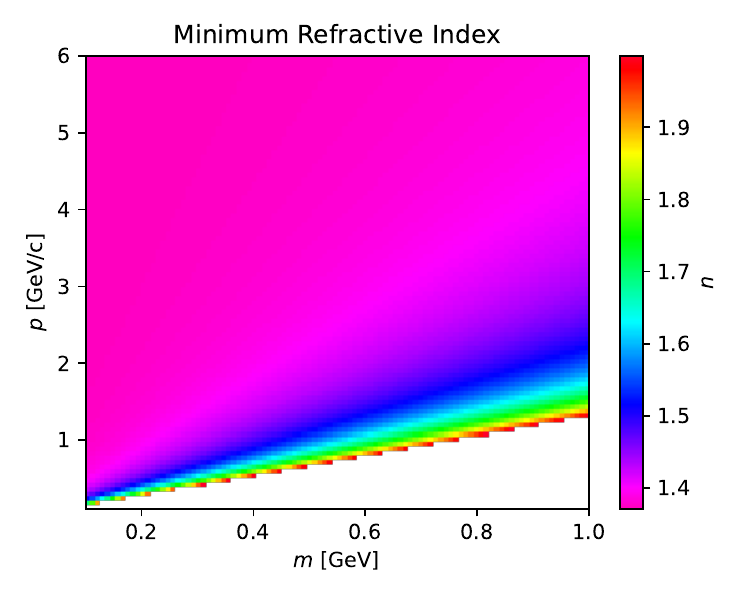}
        \caption{Refractive index as a function of mass and momentum for perpendicular charged particles.}
        \label{subfig:rindex_mass}
    \end{subfigure}
    \begin{subfigure}{\linewidth}
        \includegraphics[width=\textwidth]{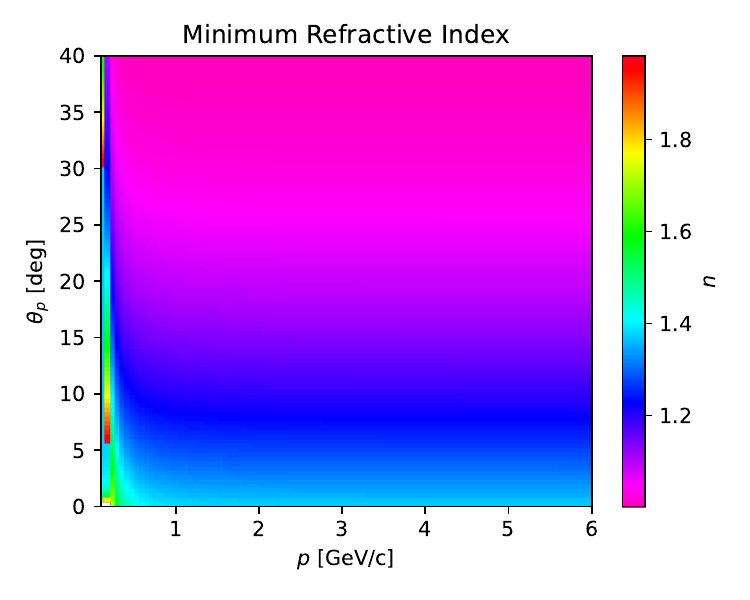}
        \caption{Refractive index for pions as a function of the polar angle of the primary particle and the momentum.}
        \label{subfig:rindex_polar_angle}
    \end{subfigure}
    \caption{Minimum required Cherenkov angle of a material for guaranteeing internal reflection of the Cherenkov photons.}
    \label{fig:rindex_mass_polar_angle}
\end{figure}

Solving equation~(\ref{eq:rindex}) for $n$ allows one to compute the minimum required refractive index as a function of the Cherenkov and particle angles.
Figure~\ref{subfig:rindex_mass} shows $n$ as a function of the primary particle's mass and momentum for a particle angle of $10^\circ$.
The threshold momentum $p_\mathrm{th}$ of a charged particle to create Cherenkov light can be directly calculated from its condition:
\begin{equation}
    v > c = \frac{c_0}{n}
\end{equation}
With $\beta = v/c = p/E$ and $E^2 = p^2 + c^2$, $p_\mathrm{th}$ is given as
\begin{equation}
    p_\mathrm{th} = \frac{m}{\sqrt{n^2 - 1}}
\end{equation}
The combinations in the histogram, showing no numbers, are below this threshold momentum.
For instance, the same plot can be created for pions as a function of the polar angle, as shown in Figure~\ref{subfig:rindex_polar_angle}.
It illustrates the increase of $n$ for larger angles, as expected.

One crucial parameter to determine the overall performance of a \gls*{dirc} detector is the number of photons trapped inside the Cherenkov radiator and the photon losses given by the inverse value.
The trapping fraction $\epsilon$ is defined as
\begin{equation}
\epsilon = \frac{N_\mathrm{trap}}{N_\mathrm{tot}}    
\end{equation}
with $N_\mathrm{trap}$ being the number of trapped photons due to internal reflection and $N_\mathrm{tot}$ being the total number of created photons by one charged particle.
The values of $\varepsilon$ depend on the refractive index for a given photon wavelength, the polar angle of the charged particle, and its momentum, which is directly related to the Cherenkov angle of the corresponding photon.

The selected material for that study is fused silica, and the assumed photon wavelength is 500\,nm.
To calculate this fraction, the photon track has to be parameterized in its polar angle $\theta_c$ and azimuth angle $\phi$ by applying spherical coordinates:
\begin{equation}
    \mathbf{r} =
    \begin{pmatrix}
    \sin\theta_c\cos\phi\\
    \sin\theta_c\sin\phi\\
    \cos\theta_c
    \end{pmatrix}
\end{equation}
However, this vector must be rotated by the charged particle's polar angle $\theta_p$.
For that purpose, it can be multiplied by the rotation matrix for a rotation around the $y$ axis:
\begin{equation}
    R = 
    \begin{pmatrix}
        \cos\theta_p & 0 & \sin\theta_p\\
        0 & 1 & 0\\
        -\sin\theta_p & 0 & \cos\theta_p
    \end{pmatrix}
\end{equation}

Because the problem is symmetric in $\phi$, the vector can be rotated around the $x$ or the $y$ axis.
Finally, the dot product of the resulting vector $\mathbf{r}' = R\mathbf{r}$ and the normal vector $\mathbf{n} = (0,0,1)$ of the radiator surface has to be computed.
As a result, the angle $\gamma$ between the photon trajectory and the normal vector can be directly calculated with the following equation
\begin{equation}
    \cos\gamma = -\sin\theta_c\cos\phi\sin\theta_p + \cos\theta_p\cos\theta_c
\end{equation}

\begin{figure}
    \centering
    \includegraphics[width=\linewidth]{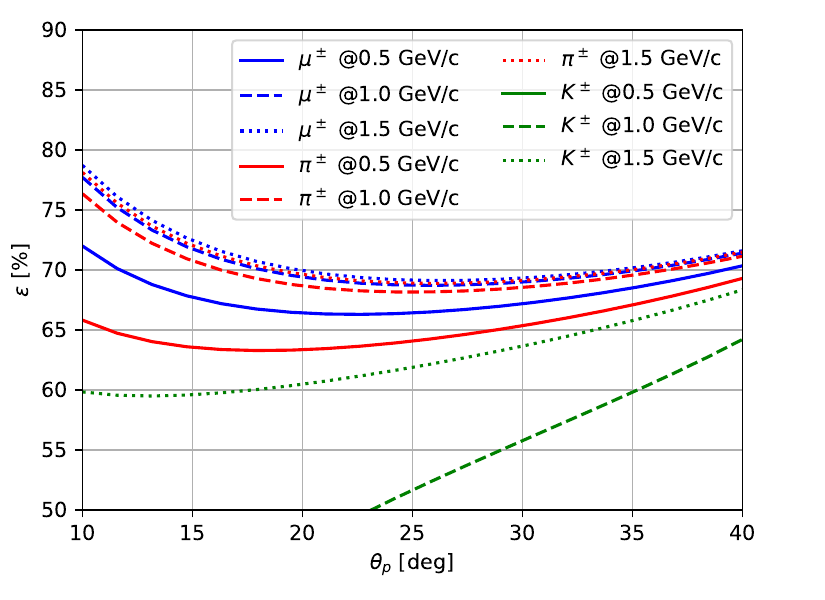}
    \caption{Trapped photons in a DIRC detector.}
    \label{fig:trapping}
\end{figure}

According to equation~(\ref{eq:internal_reflection}), the condition for internal reflection is fulfilled if
\begin{equation}
    \psi(\phi) = \gamma(\phi) - \arcsin\left(\frac{1}{n}\right) > 0
    \label{eq:condition}
\end{equation}
for a given $\phi$ angle is true.
Figure~\ref{fig:psi} has been created for all three particle species with a momentum of $p=1$\,GeV/c and a particle angle of $\theta_p = 30^\circ$.
The trapping fraction is given as the integral between $0^\circ$ and the point of intersection with the $x$-axis divided by the total area between $0^\circ$ and $180^\circ$.

\begin{figure}[!h]
    \centering
    \begin{subfigure}{\linewidth}
        \includegraphics[width=\textwidth]{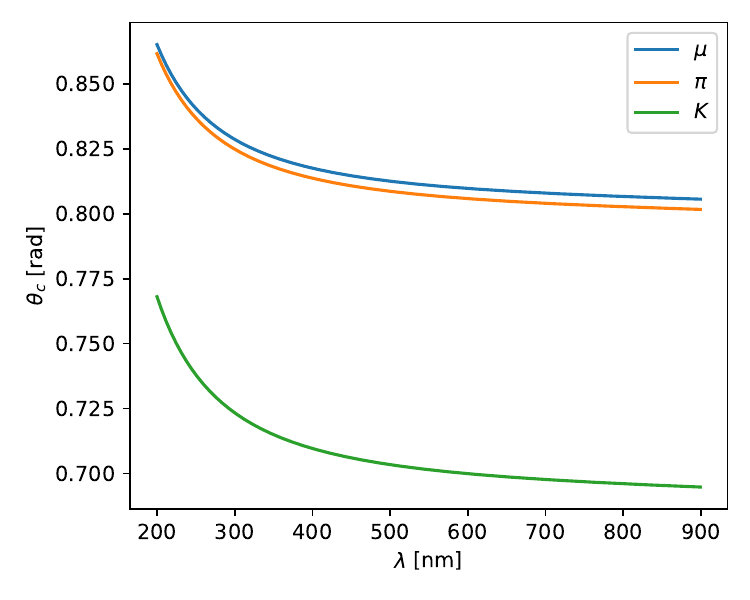}
        \caption{Cherenkov angle for three particle species.}
        \label{subfig:cherenkov}
    \end{subfigure}
    \begin{subfigure}{\linewidth}
        \includegraphics[width=\textwidth]{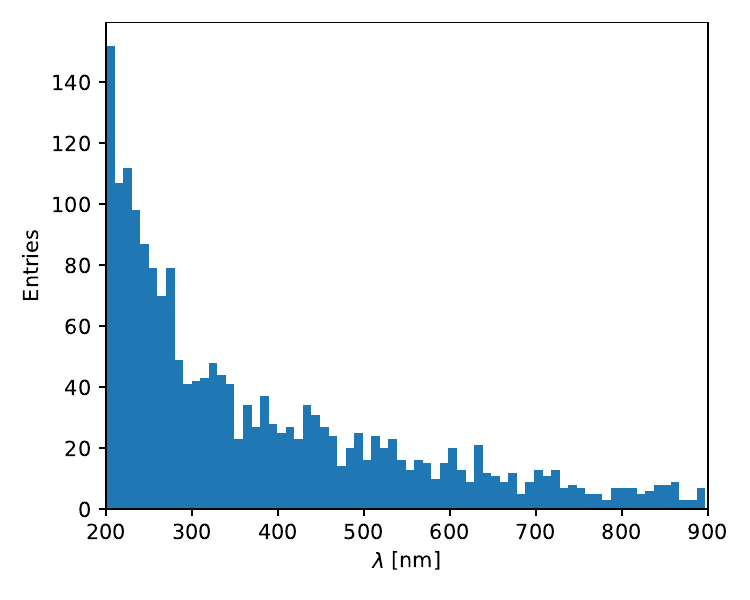}
        \caption{Wavelength distribution according to Frank-Tamm Formula.}
        \label{subfig:wavelength}
    \end{subfigure}
    \caption{Theoretically calculated Cherenkov angle as a function of the photon wavelength and the photon distribution of a muon in fused silica.}
    \label{fig:theoretical}
\end{figure}

The obtained result matches then with the trapping fraction $\epsilon$ and can be seen in Figure~\ref{fig:trapping} for three different particles and momenta.
This plot shows that the trapping fraction increases for smaller and larger particle angles and reaches a minimum between $12^\circ$ and $20^\circ$, depending on the particle species and momentum.
Smaller particle momenta result in fewer trapped photons, whereas for particles with $\ beta\rightarrow1$, the values of $\varepsilon$ converge toward a finite number as expected.
In the case of kaons at very low momenta, the condition for internal reflection is not fulfilled for low polar angles.

\section{Dispersion}

\begin{table}
    \centering
    \begin{tabular}{lcc}
    \hline
    Coefficient & Value & Unit \\
    \hline
    $B_1$ & 0.6961663 & -- \\
    $B_2$ & 0.4079426 & -- \\
    $B_3$ & 0.8974794 & -- \\
    $C_1$ & 0.0684043\textsuperscript{2} & $\mu\mathrm{m}^2$ \\
    $C_2$ & 0.1162414\textsuperscript{2} & $\mu\mathrm{m}^2$ \\
    $C_3$ & 9.896161\textsuperscript{2} & $\mu\mathrm{m}^2$ \\
    \hline
    \end{tabular}
    \caption{Sellmeier coefficients for fused silica (SiO\textsubscript{2})}
    \label{tab:sellmeier_fused_silica}
\end{table}

\begin{figure*}[!h]
    \centering
    \begin{subfigure}{0.49\textwidth}
        \includegraphics[width=\linewidth]{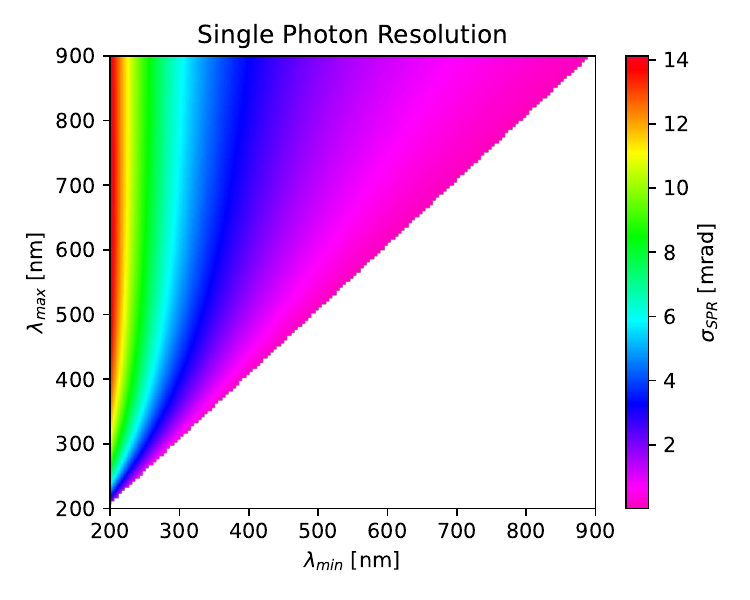}
        \caption{Single photon resolution.}
        \label{subfig:single_photon_resolution}
    \end{subfigure}
    \begin{subfigure}{0.49\textwidth}
        \includegraphics[width=\linewidth]{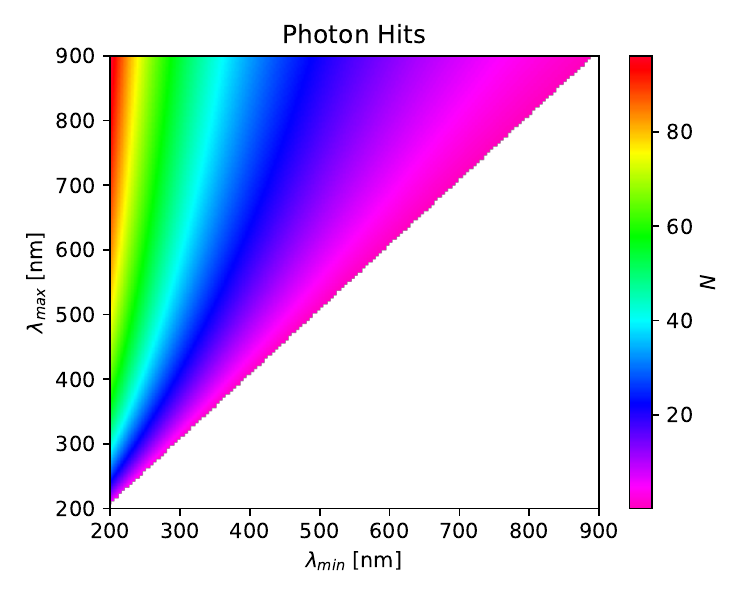}
        \caption{Number of detected photons.}
        \label{subfig:number_of_photons}
    \end{subfigure}
    \begin{subfigure}{0.49\textwidth}
        \includegraphics[width=\linewidth]{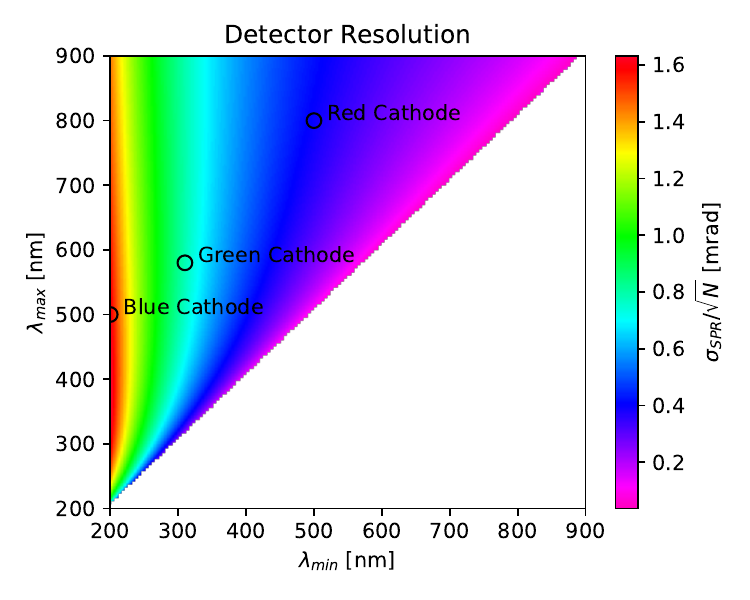}
        \caption{Total detector resolution.}
        \label{subfig:total_detector_resolution}
    \end{subfigure}
     \begin{subfigure}{0.49\textwidth}
        \includegraphics[width=\linewidth]{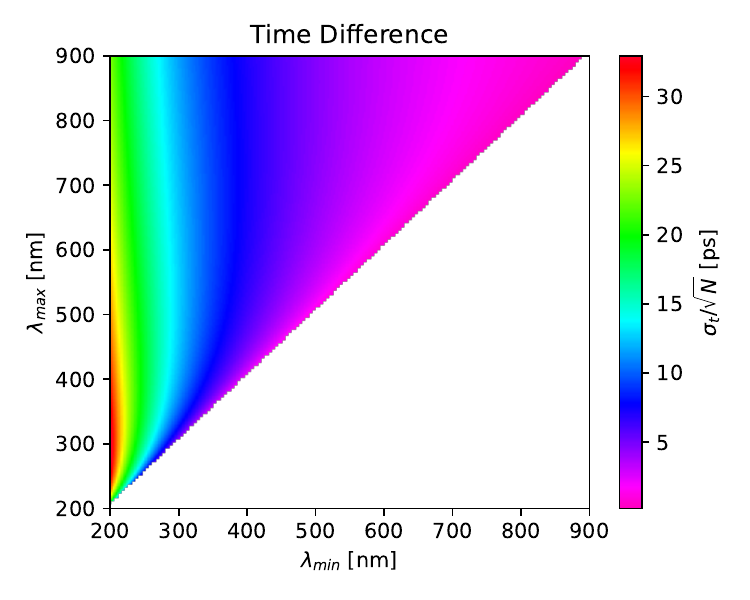}
        \caption{Time difference of photons.}
        \label{subfig:time_difference}
    \end{subfigure}
    \caption{Overall Cherenkov detector resolution resulting from dispersion in the radiator material.}
    \label{fig:dispersion_effects}
\end{figure*}

The number of created photons per track length d$x$ per wavelength interval d$\lambda$ can be calculated using the Frank-Tamm equation \cite{James2013ElectromagneticProblem}, which can be written as:
\begin{equation}
    \frac{\mathrm{d}^2N}{\mathrm{d}\lambda \mathrm{d}x} = \frac{2\pi \alpha}{\lambda^2} \left(1 - \frac{1}{\beta^2 n^2(\lambda)}\right)
    \label{eq:frank_tamm}
\end{equation}
The number of emitted Cherenkov photons depends on the inverse square root of the wavelength, i.e., the number of photons created in the blue part of the visible light is larger than in the red interval.
Furthermore, the refractive index is not a fixed number but a function of the photon wavelength.
This dependency can be written as \cite{Ghosh1997SellmeierGlasses}
\begin{equation}
    n^2(\lambda) = 1 + \sum_{i=1}^{N} \frac{B_i \lambda^2}{\lambda^2 - C_i}
    \label{eq:sellmeier}
\end{equation}
which is called the Sellmeier equation.
A two-pole version with only four coefficients is already sufficient for some materials, but for fused silica, six are required.

\begin{figure*}[!h]
    \centering
    \includegraphics[width=0.75\linewidth]{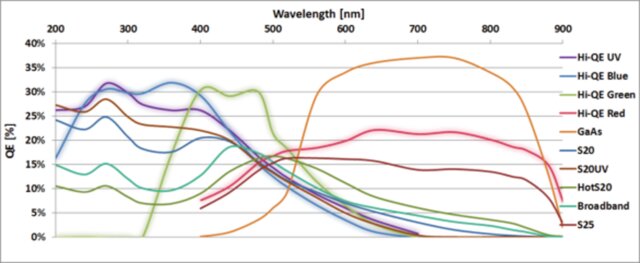}
    \caption{Quantum efficiencies for different photocathodes offered by the company Photonis \cite{Nomerotski2019ImagingCameras}.}
    \label{fig:quantum_effficiency}
\end{figure*}

The coefficients $B_i$ and $C_i$ are only defined in a specific wavelength interval (e.g., between 300\,nm and 800\,nm) and must be determined experimentally.
However, for a precise measurement, the estimated difference between the actual refractive index and the one computed with the Sellmeier equation is usually smaller than $10^{-6}$ if all six components have been determined precisely. The Sellmeier coefficients for fused silica are listed in Table~\ref{tab:sellmeier_fused_silica}.

Inserting the Sellmeier equation into equation~(\ref{eq:cherenkov_angle}) results in a wavelength dependency of the Cherenkov angle as shown in Figure~\ref{subfig:cherenkov} for three different particles with momenta of 1\,GeV/c in the wavelength window of 200 to 900\,nm.
The steep curve fall at smaller wavelengths results in a larger $\sigma_\mathrm{disp}$ value.
Figure~\ref{subfig:wavelength} illustrates the photon distribution for a muon with a momentum of 1\,GeV passing through a fused silica radiator with a thickness of 2\,cm.
This result is directly obtained from the Frank-Tamm equation by first integrating this formula with respect to $\lambda$ to receive the total number of photons, around 1,950.
In the next step, uniformly distributed random numbers were generated, and the rejection method was used to decide whether a photon should be created for this wavelength interval.
Each histogram bin contains the number of photons within a wavelength interval of 10\,nm.

All dispersion-related resolution studies were done by applying a minimum and maximum range cut for the created photon wavelengths.
In Figure~\ref{subfig:single_photon_resolution}, the single photon resolution can be found as a function of the wavelength interval boundaries.
If the minimum wavelength is given as 200\,nm and the maximum as 900\,nm, the Cherenkov angle smearing reaches its maximum around 14\,mrad.
As expected, the resolution improves by shrinking the interval to accept only large wavelengths, up to values below 1\,mrad.

However, the number of detected photons decreases when narrowing the wavelength interval, as shown in Figure~\ref{subfig:number_of_photons}.
For this calculation, a radiator thickness of 2\,cm was assumed, as the number of created photons scales linearly with the particle path length inside the medium at first order.
To obtain realistic results, an overall photon loss of 70\%, as taken from Figure~\ref{fig:trapping}, was further assumed in addition to an average detection efficiency of 20\%.
The photon numbers match with other studies of \gls*{dirc} detectors, such as the \gls*{panda} Barrel or Disc \gls*{dirc} counters as well as measurements of various photon sensors \cite{Lehmann2020RecentPMTs}.
As a result, the number of detected photons varies between 0 and 90 depending on the chosen window.

And finally, the overall detector resolution can be calculated by dividing the single photon resolution by the square root of the detected photons, resulting in Figure~\ref{subfig:total_detector_resolution}.
The results indicate the best performance of a detector at a narrow wavelength window around large wavelengths.
For larger windows and smaller wavelengths, the resolution deteriorates significantly.
This effect results from the non-linear scaling of the resolution to the number of photons.

The question arises if the Cherenkov angle smearing can be mitigated by measuring the \gls*{top} with high precision.
The resulting plot is presented in Figure~\ref{subfig:time_difference} showing the time resolution as a function of the different wavelength windows divided by the square root of the number of detected photons.
The calculation was done similar to the one for obtaining the spatial resolution but using equations~(\ref{eq:time}) and (\ref{eq:distance}).
The largest value is located around 250\,ps for short windows containing short wavelengths.
Especially for $\mu/\pi$ separation, the calculated standard deviation exceeds the required detector resolution shown in Figure~\ref{subfig:time_resolution_mupi}.

The labels for blue, green, and red photocathodes refer to a datasheet published by the company Photonis  \cite{Nomerotski2019ImagingCameras} (as shown in Figure~\ref{fig:quantum_effficiency}), a leading manufacturer of \gls*{mcp}-\glspl*{pmt}.
Generally, the cathodes with a high quantum efficiency in the red spectrum result in the best detector resolution.
An ideal candidate would be the gallium arsenide cathode with more than 30\% quantum efficiency in the red spectrum.
However, these cathodes' lifetime and radiation hardness are not sufficient for use in particle physics experiments.
The green photocathode with an enhanced quantum efficiency in the green spectrum can be considered a good compromise.
Still, the results would not guarantee a sufficient separation of muons and pions at larger momenta.
Thicker radiators result in a larger photon yield but increase the smearing induced by angle-straggling.
Finding an optimum value is crucial for a sufficient $\mu/\pi$ separation.

\section{Sensor Granularity}

\begin{figure}[!h]
    \centering
    \includegraphics[width=\linewidth]{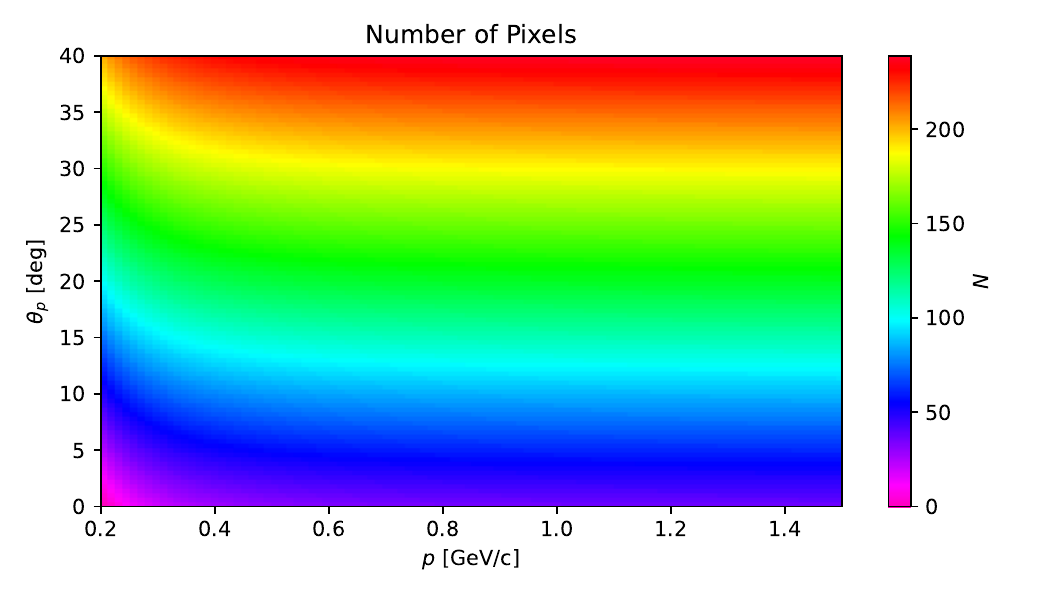}
    \caption{Number of required pixels for achieving a desired resolution of 1\,mrad with muons.}
    \label{fig:pixels}
\end{figure}

The number of pixels for photon detection is crucial in optimizing the overall detector resolution.
The standard deviation of a uniform distribution with the limits $a$ and $b$ is given as
\begin{eqnarray}
    \sigma = \sqrt{\overline{x^2} - \overline{x}^2} = \frac{1}{b-a}\int_a^b x^2\,\mathrm{d}x -\\ \left(\frac{1}{b-a}\int_a^b x\,\mathrm{d}x\right)^2 = \frac{b-a}{\sqrt{12}}
\end{eqnarray}
which is identical to the pixel width divided by the factor $\sqrt{12}\approx 3.5$.
Based on this correlation, Figure~\ref{fig:pixels} shows the required pixels to achieve a 1\ mrad resolution with muons as a function of the particle's polar angle $\theta_p$ and its momentum $p$.
The particle momentum and polar angle interval is the distance between the displayed and minimum values.
The values increase with the maximum value of the polar angle as expected.
The number of pixels converges to a finite number concerning the particle momentum very quickly.
For $\theta_p = 40^\circ$ and $\beta\rightarrow 1$, up to 250 pixels would be required, starting from $\theta = 0^\circ$ and $p = 0.2$\,GeV/c.
For the \gls*{edd} in \gls*{panda}, an angle window $21^\circ < \varphi \leq 42^\circ$ was foreseen, which required a total number of 100 pixels in $\theta$ direction.

\section{Angle Straggling}

\begin{figure}[!h]
    \centering
    \includegraphics[width=\linewidth]{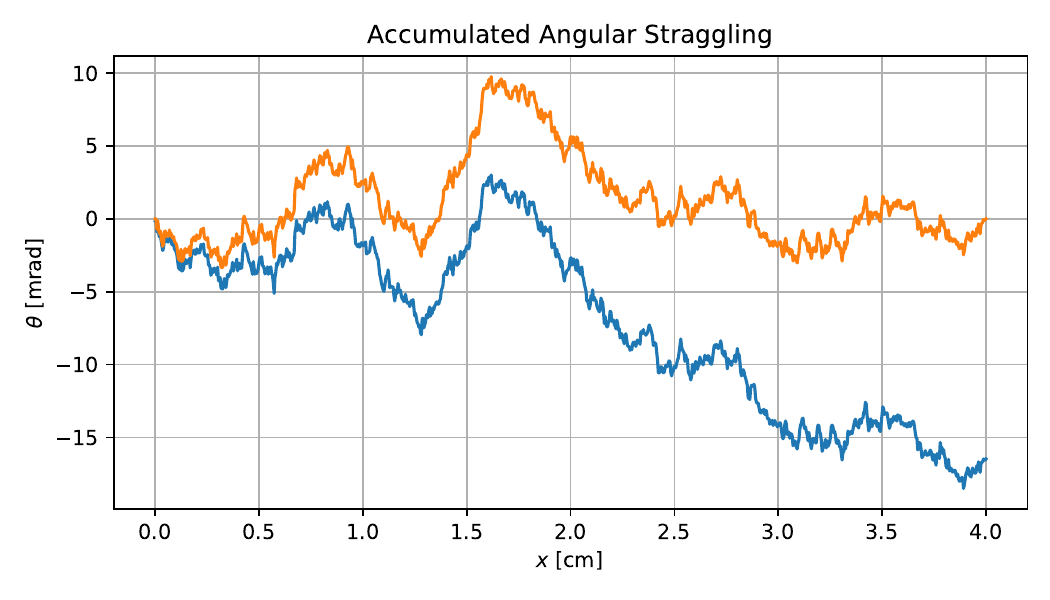}
    \caption{Path of a muon (blue) with a momentum of 0.5\,GeV/c inside a radiator with a thickness of 4\,cm and the corrected path (orange) by taking the initial and final angle into account.}
    \label{fig:straggling}
\end{figure}

For calculating the radiation length of a material with the number of protons $Z$ and nucleons $A$, the following formula can be used \cite{Lynch1991ApproximationsScattering}:
\begin{equation}
    \frac{1}{X_0} = 4 \alpha r_e^2 N_A \frac{Z^2}{A} \left[ \ln\left(\frac{183}{Z^{1/3}}\right) \right]
    \label{eq:inverse_radiation_length}
\end{equation}
with the fine structure constant $\alpha$, the electron radius $r_e$, and the Avogadro constant $N_A$.
For a compound material consisting of multiple elements, the sum over each element's radiation length must be computed, including the weight factor $w_i$ for each element:
\begin{equation}
  \frac{1}{X_0} = \sum_i \frac{w_i}{X_0(i)}
  \label{eq:radiation_length}
\end{equation}
With the radiation length, the \gls*{rms} of the particle's polar angle after passing a distance of $x$ inside a medium with the thickness $x$ can be calculated according to \cite{Zyla2020ReviewPhysics}:
\begin{equation}
    \theta_0 = \frac{13.6 \, \text{MeV}}{\beta c p} z \sqrt{\frac{x}{X_0}} \left[1 + 0.038 \ln\left(\frac{x}{X_0}\right)\right]
    \label{eq:highland_formula}
\end{equation}
This equation was used to calculate random walks of muons and pions through a fused silica radiator semi-analytically.
An example for a muon with a momentum of 1.5\,GeV/c and a radiator thickness of 4\,cm can be found in Figure~\ref{fig:straggling}.
The particle entered perpendicular to the radiator's surface, which was divided into 1000 slices.

In every slice, the new angle $\theta'$ is calculated using a Gaussian distribution according to
\begin{equation}
    \theta_\mathrm{new} = \frac{1}{\sqrt{2\pi}\theta_0} \exp\left[-\frac{1}{2}\left(\frac{\theta_\mathrm{old} - \theta}{\theta_0}\right)^2\right]
\end{equation}
At the end, the \gls{rms} value
\begin{equation}
    \sigma_{\theta} = \sqrt{\frac{1}{n}\sum x_i^2}
\end{equation}
of all scattering angles was computed, which is identical to the Cherenkov angle smearing due to angle straggling because the Cherenkov photons are uniformly emitted along the particle track, and the mean value is unknown.

\begin{figure}[!h]
    \centering
    \begin{subfigure}{\linewidth}
        \includegraphics[width=\textwidth]{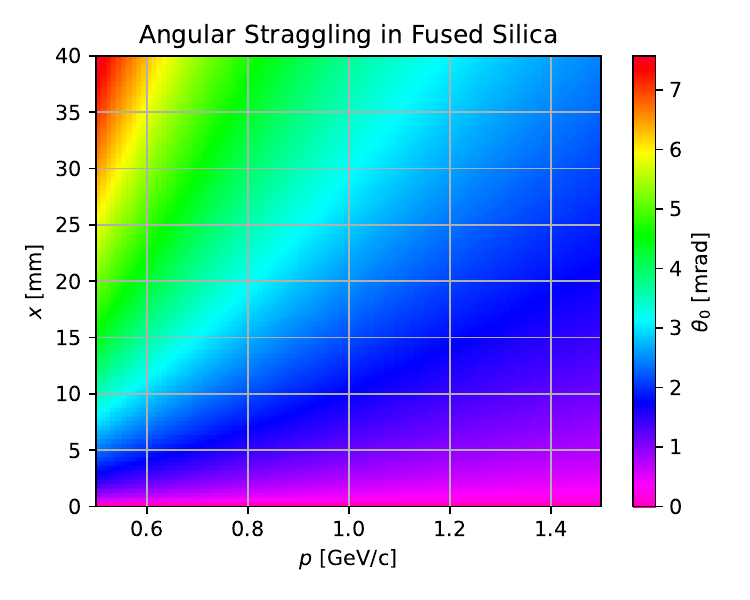}
        \caption{Angle straggling for $\mu^\pm$.}
        \label{subfig:straggling_muon_uncorrected_step}
    \end{subfigure}
    \begin{subfigure}{\linewidth}
        \includegraphics[width=\textwidth]{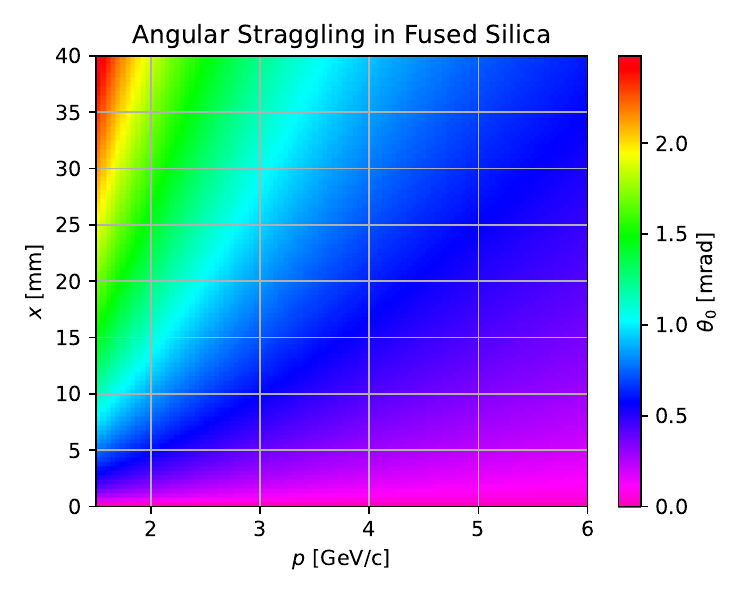}
        \caption{Angle straggling for $\pi^\pm$.}
        \label{subfig:straggling_pion_uncorrected_step}
    \end{subfigure}
    \caption{Cherenkov angle smearing resulting from multiple scattering due to Coulomb interactions of the charged particle with the radiator.}
    \label{fig:straggling_uncorrected}
\end{figure}

Figure~\ref{subfig:straggling_muon_uncorrected_step} shows the results for muons in the momentum range between 0.5 and 1.5\,GeV/c, whereas Figure~\ref{subfig:straggling_pion_uncorrected_step} displays the results for pions from 0.5 to 6\,GeV/c.
The largest value of 7.2\,mrad is obtained for muns at 0.5\,GeV/c momentum for a radiator thickness of 4\,cm.
For a thickness of 2\,cm and a momentum of 1.5\,GeV/c the smearing is still 1.6\,mrad which is greater than the required resolution.
It turns out that adding another tracking layer on the bottom of the surface and measuring the particle's momentum direction at this position can improve the resolution by a factor of $\sqrt{3}$.

\begin{figure}[!h]
    \centering
    \begin{subfigure}{0.4\textwidth}
        \includegraphics[width=\textwidth]{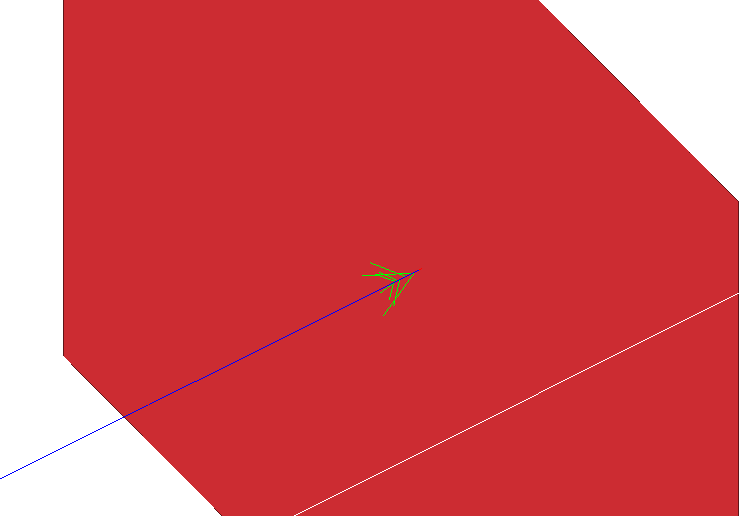}
        \caption{Geant4 event display.}
        \label{subfig:geant4_event_display}
    \end{subfigure}
    \quad
    \begin{subfigure}{\linewidth}
        \includegraphics[width=\textwidth]{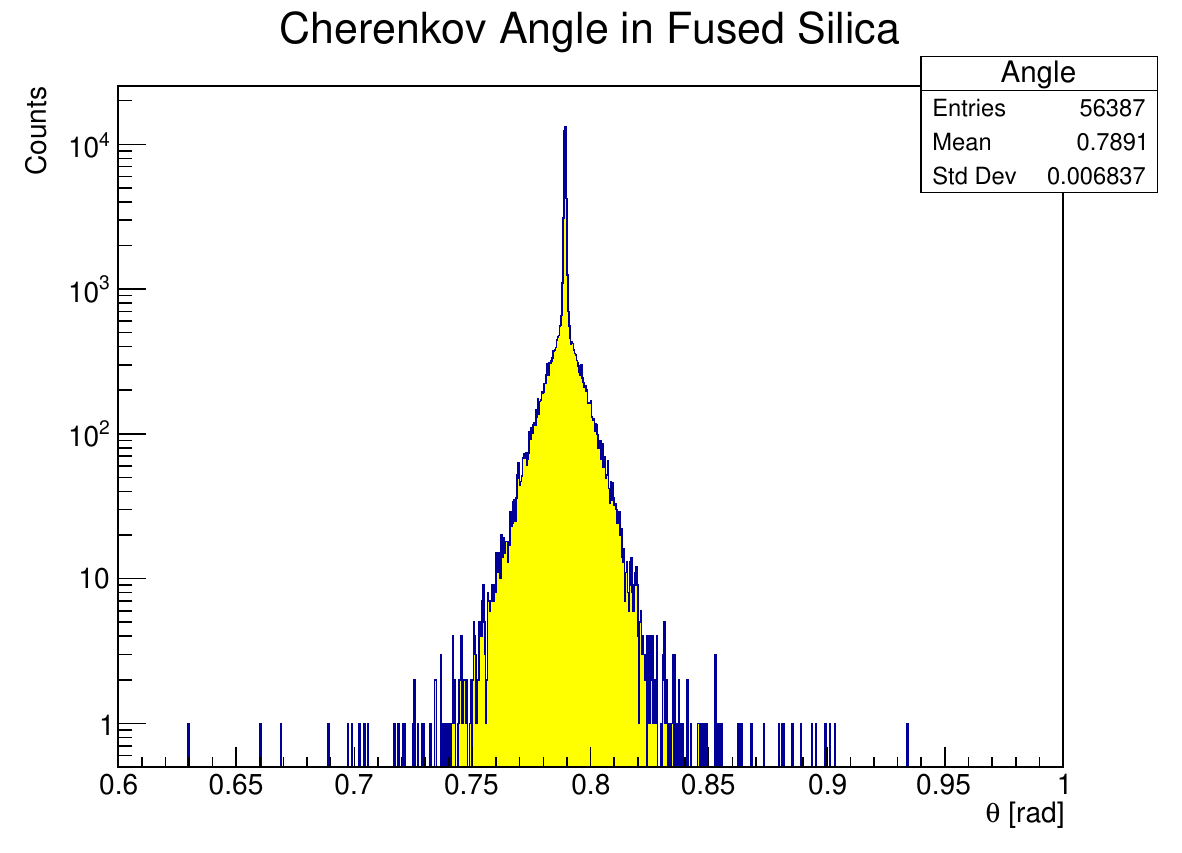}
        \caption{Histogram of Cherenkov angles.}
    \end{subfigure}
    \caption{Results obtained with full Geant4 detector simulation.}
    \label{subfig:geant4_straggling}
\end{figure}

The validity of the results was verified for some data points using a Geant4 detector simulation, showing an excellent match, especially for larger momenta and radiator thicknesses.
Figure~\ref{subfig:geant4_event_display} shows the event display of the simulated muon tracks with a momentum of 0.5\,GeV inside a fused silica radiator with a thickness of 4\,cm.
The properties of fused silica are directly taken from the NIST material database of Geant4.
The material's refractive index as a function of the photon wavelength was calculated using Sellmeier's equation (\ref{eq:sellmeier}).
However, the wavelength interval was chosen to 10\,nm at large wavelengths to ensure that the smearing due to dispersion is reduced to a minimum.
The simulation of a larger number of particle tracks compensated for the smaller number of photons.
For each simulation step of the primary particle, the global polar angle $\theta$ of the created photon is filled into a histogram to obtain the resulting \gls*{rms} value as shown in Figure~\ref{subfig:geant4_straggling} for a total of $10^4$ simulated muon tracks.

\section{Energy Loss}

Assuming that the energy loss of a charged particle inside matter is only a result of ionization and radiation losses can be ignored, the Bethe-Bloch formula can be used to calculate the average value of d$E$/d$x$:
\begin{eqnarray}
    -\left\langle \frac{\mathrm{d}E}{\mathrm{d}x} \right\rangle = K \cdot z^2 \cdot \frac{Z}{A} \cdot \frac{1}{\beta^2} \cdot \\
    \left[ \frac{1}{2} \ln\left(\frac{2m_e c^2 \beta^2 \gamma^2 T_{\text{max}}}{I^2}\right) - \beta^2 - \frac{\delta}{2} \right]
    \label{eq:bethe_bloch}
\end{eqnarray}
with $T_\mathrm{max}$ given as:
\begin{equation}
    T_{\text{max}} = \frac{2 m_e c^2 \beta^2 \gamma^2}{1 + 2 \gamma \frac{m_e}{m} + \left( \frac{m_e}{m} \right)^2}
    \label{eq:tmax_bethe_bloch}
\end{equation}
For the effective $Z/A$ ratio, a value of 0.5 for fused silica is used widely in many publications \cite{Torres-Torres2010GeometrySimulation}.
The mean excitation energy $I$ can be calculated can be approximated very well with $I = 10\,\mathrm{eV} Z$.
However, for fused silica, we have to insert this value into the following formula for the weighted average:
\begin{equation}
    \ln I = \frac{\sum w_i \cdot \frac{Z_i}{A_i} \cdot \ln I_i}{\sum w_i \cdot \frac{Z_i}{A_i}}
\end{equation}
which results in a value of 125\,eV.
This value is close to the one commonly used in research, given as 135\,eV \cite{Bull1986StoppingPositrons}.
The last term $\delta/2$ of the Bethe-Bloch formula related to density corrections has been ignored for the following calculations.

\begin{figure}[!h]
    \centering
    \begin{subfigure}{\linewidth}
        \includegraphics[width=\textwidth]{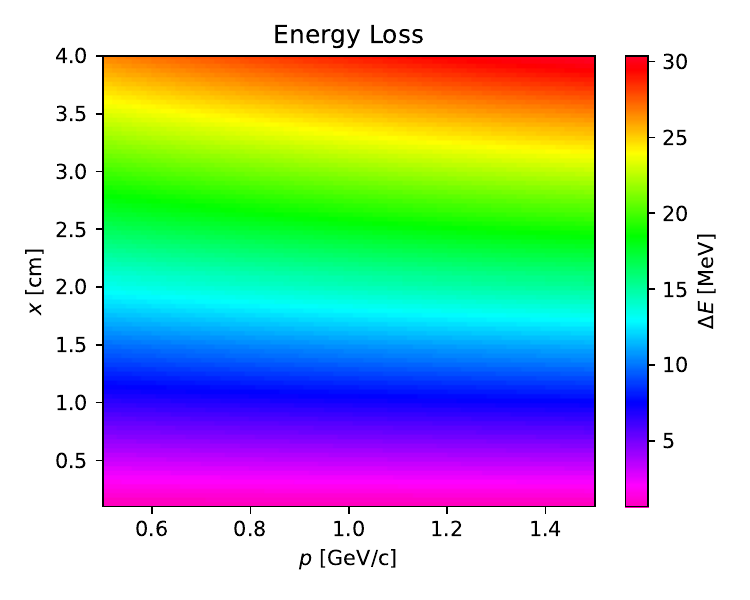}
        \caption{Energy loss.}
        \label{subfig:energy_loss}
    \end{subfigure}
    \begin{subfigure}{\linewidth}
        \includegraphics[width=\textwidth]{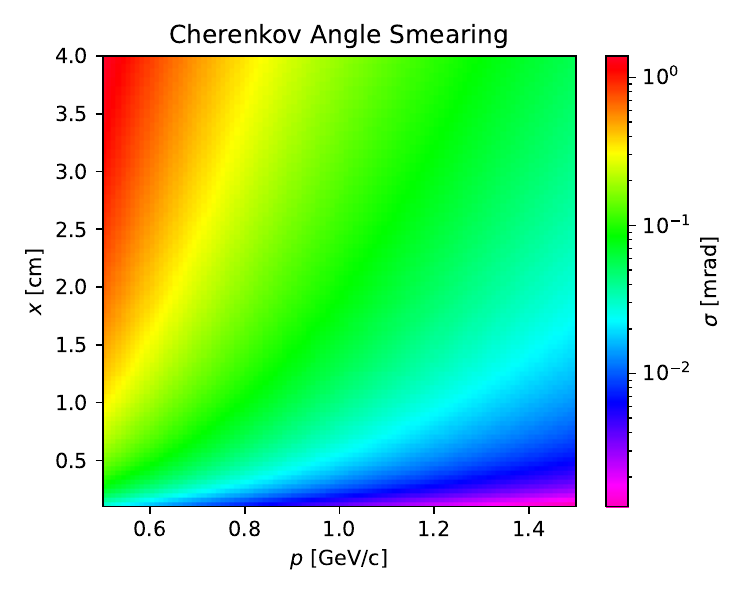}
        \caption{Cherenkov angle smearing.}
        \label{subfig:cherenkov_energy}
    \end{subfigure}
    \caption{Energy loss and smearing of Cherenkov angle of muons as a function of the initial momentum and radiator thickness due to energy loss in fused silica.}
    \label{subfig:energy_loss_cherenkov_smearing}
\end{figure}

A numerical integration of the Bethe-Bloch equation was done by dividing the radiator into thin slices with 0.1\,mm thickness each.
The total energy loss as a function of the radiator thickness and initial particle momentum can be seen in Figure~\ref{subfig:energy_loss}.
The largest value given as 30\,MeV was obtained for a thickness of 4\,cm and a momentum of 1.5\,GeV/c.
The related Cherenkov angle smearing has been computed directly using equation~(\ref{eq:cherenkov_angle}) with $p$ being the remaining particle momentum at each integration step.
From all steps, the \gls*{rms} value was calculated similarly to the smearing related to angle straggling.
The results are illustrated in Figure~\ref{subfig:cherenkov_energy} with a logarithmic color axis.
The values differ from $10^{-3}$\,mrad for thin radiators with large particle momenta up to 1.5\,mrad at a momentum of 0.5\,GeV/c and 4\,cm thickness.

\section{Conclusion}

This paper systematically analyzes the resolution limits of all components of \gls*{dirc} detectors, focusing on photon loss, dispersion effects, and angle straggling.
The studies demonstrate that while some deterioration, such as sensor granularity and surface roughness, can be minimized through optimization processes, fundamental limitations remain due to the physical nature of Cherenkov photon emission and particle interactions within the radiator.

The calculations confirm that dispersion imposes a strong wavelength-dependent smearing on the Cherenkov angle, which can be partially mitigated by wavelength filters or the sensor's quantum efficiency range.
However, this comes at the cost of photon statistics.
Similarly, angle straggling significantly affects low-momentum particles and imposes an intrinsic limit on the resolution that cannot be improved by increasing the number of detected photons.
Additional tracking stations can increase the resolution, but the resolution of these tracking detectors must also be considered.
Furthermore, energy loss effects are subdominant but become relevant for thick radiators and low-energy tracks.

The trapping efficiency study further underlines the importance of optimizing the geometry and material properties to enhance internal reflection and photon yield.
The presented results provide valuable input for the design and feasibility evaluation of future \gls*{pid} systems, particularly those aiming for high-resolution performance in the low-to-intermediate-momentum range.
Ultimately, it turns out that finding the best setup for a \gls*{dirc} detector results in a multi-parameter optimization and has to be studied precisely.

More comprehensive Geant4 simulations, including whole detector geometry and readout electronics, could validate and refine the analytical models used in this study.
Furthermore, future developments of dispersion correction could reduce the resolution limitations resulting from dispersion effects.
Finally, the applicability of these findings to dedicated experiments beyond \gls*{panda}, \gls*{sctf}, and \gls*{epic} can be further studied in more detail, especially in the context of compact and cost-efficient \gls*{dirc} designs.

\bibliographystyle{JHEP}
\bibliography{biblio}

\providecommand{\href}[2]{#2}\begingroup\raggedright\begin{thebibliography}{10}

\bibitem{Belias2023OverviewFAIR}
A.~Belias, \emph{{Overview of the PANDA detector design at FAIR}}, \href{https://doi.org/10.1142/s2010194523600017}{\emph{International Journal of Modern Physics: Conference Series} {\bfseries 51} (2023) }.

\bibitem{Piminov2018ProjectBINP}
P.~Piminov, \emph{{Project for a Super Charm–Tau Factory at BINP}}, \href{https://doi.org/10.1134/S1547477118070579}{\emph{Physics of Particles and Nuclei Letters} {\bfseries 15} (2018) }.

\bibitem{Allaire2024ArtificialAI4EIC}
C.~Allaire, R.~Ammendola, E.C.~Aschenauer, M.~Balandat, M.~Battaglieri, J.~Bernauer et~al., \emph{{Artificial Intelligence for the Electron Ion Collider (AI4EIC)}},  2024.
\newblock 10.1007/s41781-024-00113-4.

\bibitem{Singh2019TechnicalDetector}
B.~Singh, W.~Erni, B.~Krusche, M.~Steinacher, N.~Walford, B.~Liu et~al., \emph{{Technical design report for the PANDA Barrel DIRC detector}}, \href{https://doi.org/10.1088/1361-6471/aade3d}{\emph{Journal of Physics G: Nuclear and Particle Physics} {\bfseries 46} (2019) }.

\bibitem{Davi2022TechnicalDIRC}
F.~Dav{\`{i}}, W.~Erni, B.~Krusche, M.~Steinacher, N.~Walford, H.~Liu et~al., \emph{{Technical design report for the endcap disc DIRC}}, \href{https://doi.org/10.1088/1361-6471/abb6c1}{\emph{Journal of Physics G: Nuclear and Particle Physics} {\bfseries 49} (2022) }.

\bibitem{Miehling2023LifetimePhotomultipliers}
D.~Miehling, M.~B{\"{o}}hm, K.~Gumbert, S.~Krauss, A.~Lehmann, A.~Belias et~al., \emph{{Lifetime and performance of the very latest microchannel-plate photomultipliers}}, \href{https://doi.org/10.1016/j.nima.2023.168047}{\emph{Nuclear Instruments and Methods in Physics Research, Section A: Accelerators, Spectrometers, Detectors and Associated Equipment} {\bfseries 1049} (2023) }.

\bibitem{Barnyakov2023CalibrationFactory}
A.Y.~Barnyakov, V.S.~Bobrovnikov, S.A.~Kononov, P.D.~Rogozhin and M.V.~Chadeeva, \emph{{Calibration and Reconstruction Algorithm for FARICH System of the Detector at Super Charm-Tau Factory}}, \href{https://doi.org/10.3103/S1068335623120047}{\emph{Bulletin of the Lebedev Physics Institute} {\bfseries 50} (2023) }.

\bibitem{Hayrapetyan2022PlansNovosibirsk}
A.~Hayrapetyan, S.~Kegel, M.~Schmidt, A.Y.~Barnyakov, V.S.~Bobrovnikov and S.A.~Kononov, \emph{{Plans for novel Cherenkov detectors at the Super Charm-Tau Factory in Novosibirsk}},  in \emph{Journal of Physics: Conference Series}, vol.~2374, 2022, \href{https://doi.org/10.1088/1742-6596/2374/1/012122}{DOI}.

\bibitem{Kalicy2024TheEIC}
G.~Kalicy, \emph{{The high-performance DIRC for the ePIC detector at the EIC}}, \href{https://doi.org/10.1016/j.nima.2024.169168}{\emph{Nuclear Instruments and Methods in Physics Research, Section A: Accelerators, Spectrometers, Detectors and Associated Equipment} {\bfseries 1062} (2024) }.

\bibitem{Wu2008TextbookI}
M.~Wu, X.~Fan, Q.~Liu, Z.~He, A.~Mateeva, J.~Lopez et~al., \emph{{Textbook Of Engineering Physics (Part I)}}, {\emph{Applied Optics} {\bfseries 62} (2008) }.

\bibitem{Born2000PrinciplesWolf}
M.~Born and E.~Wolf, \emph{{Principles of Optics M. Born and E. Wolf}},  2000.

\bibitem{Wray1969RefractiveTemperature}
J.H.~Wray and J.T.~Neu, \emph{{Refractive Index of Several Glasses as a Function of Wavelength and Temperature*}}, \href{https://doi.org/10.1364/josa.59.000774}{\emph{Journal of the Optical Society of America} {\bfseries 59} (1969) }.

\bibitem{Bartell1981titleTheBTDF/title}
F.O.~Bartell, E.L.~Dereniak and W.L.~Wolfe, \emph{{<title>The Theory And Measurement Of Bidirectional Reflectance Distribution Function (Brdf) And Bidirectional Transmittance Distribution Function (BTDF)</title>}},  in \emph{Radiation Scattering in Optical Systems}, vol.~0257, 1981, \href{https://doi.org/10.1117/12.959611}{DOI}.

\bibitem{Kovalenko2001Descartes-SnellAbsorption}
S.~Kovalenko, \emph{{Descartes-Snell law of refraction with absorption}}, \href{https://doi.org/10.15407/spqeo4.03.214}{\emph{Semiconductor Physics, Quantum Electronics and Optoelectronics} {\bfseries 4} (2001) }.

\bibitem{James2013ElectromagneticProblem}
C.W.~James, \emph{{Electromagnetic radiation in the Tamm problem}},  in \emph{AIP Conference Proceedings}, vol.~1535, 2013, \href{https://doi.org/10.1063/1.4807539}{DOI}.

\bibitem{Ghosh1997SellmeierGlasses}
G.~Ghosh, \emph{{Sellmeier coefficients and dispersion of thermo-optic coefficients for some optical glasses}}, \href{https://doi.org/10.1364/ao.36.001540}{\emph{Applied Optics} {\bfseries 36} (1997) }.

\bibitem{Lehmann2020RecentPMTs}
A.~Lehmann, M.~B{\"{o}}hm, D.~Miehling, M.~Pfaffinger, S.~Stelter, F.~Uhlig et~al., \emph{{Recent progress with microchannel-plate PMTs}}, \href{https://doi.org/10.1016/j.nima.2019.01.047}{\emph{Nuclear Instruments and Methods in Physics Research, Section A: Accelerators, Spectrometers, Detectors and Associated Equipment} {\bfseries 952} (2020) }.

\bibitem{Nomerotski2019ImagingCameras}
A.~Nomerotski, \emph{{Imaging and time stamping of photons with nanosecond resolution in Timepix based optical cameras}},  2019.
\newblock 10.1016/j.nima.2019.05.034.

\bibitem{Lynch1991ApproximationsScattering}
G.R.~Lynch and O.I.~Dahl, \emph{{Approximations to multiple Coulomb scattering}}, \href{https://doi.org/10.1016/0168-583X(91)95671-Y}{\emph{Nuclear Inst. and Methods in Physics Research, B} {\bfseries 58} (1991) }.

\bibitem{Zyla2020ReviewPhysics}
P.A.~Zyla, R.M.~Barnett, J.~Beringer, O.~Dahl, D.A.~Dwyer, D.E.~Groom et~al., \emph{{Review of particle physics}},  2020.
\newblock 10.1093/ptep/ptaa104.

\bibitem{Torres-Torres2010GeometrySimulation}
D.~Torres-Torres, J.~Mu{\~{n}}oz-Salda{\~{n}}a, L.A.L.~De~Guevara, A.~Hurtado-Mac{\'{i}}as and M.V.~Swain, \emph{{Geometry and bluntness tip effects on elastic-plastic behaviour during nanoindentation of fused silica: Experimental and FE simulation}}, \href{https://doi.org/10.1088/0965-0393/18/7/075006}{\emph{Modelling and Simulation in Materials Science and Engineering} {\bfseries 18} (2010) }.

\bibitem{Bull1986StoppingPositrons}
R.~Bull, \emph{{Stopping powers for electrons and positrons}}, \href{https://doi.org/10.1016/1359-0189(86)90048-8}{\emph{International Journal of Radiation Applications and Instrumentation. Part D. Nuclear Tracks and Radiation Measurements} {\bfseries 11} (1986) }.

\end{thebibliography}\endgroup

\end{document}